\documentclass[a4paper, twocolumn, superscriptaddress,prl]{revtex4}  %% REVTeX 4.0
\usepackage{graphicx}
\usepackage{amsmath,amssymb}
\begin{document}
\title{Fluorescence antibunching microscopy}
\author{O. Schwartz}
\author{D. Oron}
\affiliation{Department of Physics of Complex Systems, Weizmann Institute of
Science, Rehovot, Israel}
\begin{abstract}
Breaking the diffraction limit in microscopy by utilizing quantum properties of
light has been the goal of intense research in the recent years. We propose a
quantum superresolution technique based on non-classical emission statistics of
fluorescent markers, routinely used as contrast labels for bio-imaging. The
technique can be readily implemented using standard fluorescence microscopy
equipment.
\end{abstract}
\maketitle Increasing the imaging resolution in optical microscopy can
potentially benefit many fields of research, including life sciences. In
classical linear optics, diffraction imposes a limit on the resolution of
far-field microscopy. In the last two decades, a number of techniques have been
developed that break this limit by making use of nonlinear optical
processes\cite{STED_Hell_OL1994, KawataPRL07, Gustafsson_SSI_PNAS05}. The
diffraction limit can also be effectively overcome by utilizing contrast labels
exhibiting strong variations in brightness, either induced by
photoactivation\cite{PALM_Betzig_Science2006, PALM_Sam_Hess_BiophysJ2006,
STORM_NatMet06} or intrinsic
\cite{Localization_by_blinking_Heintzmann_OpEx2005, SOFI_DertingerPNAS2009}, in
effect allowing for localization of individual fluorophores while the rest are
dark.

Quantum optics offers another promising pathway to superresolution imaging.
Quantum optical methods have been shown to dramatically increase the resolution
in interferometric
measurements\cite{Quantum_Enhanced_measurements_Science2004}, and allowed for
imaging sensitivity enhancement beyond shot
noise\cite{Sub_shot_noise_imaging_NatPhot2010}. At the same time,
superresolution via quantum imaging has not yet been demonstrated
experimentally.
The theoretical research aiming at achieving sub-diffraction limited quantum
imaging has mainly focused on the scenario wherein an absorptive object is
illuminated with a non-classical light beam. Superresolution is then attained
via two main routes. One requires an object consisting of or stained with a
multi-photon absorbing material. In this scheme, the diffraction limit is
overcome by utilizing the high spatial frequency quantum interference patterns,
similarly to quantum superresolution
lithography\cite{Lithography_proof_of_principle_PRL2001}. Although multi-photon
interference patterns of high order have been observed experimentally using
coincidence detection\cite{Afek_Science2010,
Four_photon_Interference_Zeilinger_Nature2004}, the lack of a low light level
multi-photon absorber makes this approach currently not feasible.
The other pathway to achieving sub-diffraction resolution is probing an object
with a beam of light exhibiting position (or momentum) entanglement. The high
resolution images can then be obtained simply by coincidence
detection\cite{Giovannetti_Number_Resolving_PRA2009, Centroids_PRL2009}.
Although momentum entangled light can be produced by spontaneous parametric
down conversion\cite{deBroglie_Wavelength_Fonseca_PRL1999,
Spatial_Antibunching_Monken_PRL2001}, the resolution in this case is limited by
the diffraction limit at the pumping wavelength. Increasing the resolution
further requires a bright highly entangled light source, which is yet to be
developed.

The above superresolution schemes exploit quantum properties of the
illuminating light and thus require non-classical light sources. The
alternative is, to image an object naturally emitting non-classical light, such
as cascaded emission from a three-level system
\cite{Quantum_microscopy_using_photon_correlations_Muthukrishnan_JOB2004} or
resonant fluorescence\cite{Incoherent_PRL2007}. The multi-photon interference
patterns observable in the far field can be used for superresolved imaging of
the emitters. While this approach can be feasible for imaging trapped ions, its
reliance on fragile quantum effects makes it impractical for bio-imaging
conditions.

In this paper, we consider a different property of fluorescence emitters:
photon antibunching\cite{Antibunching_Kimble_PRL1977}, arising from the
tendency of fluorophores to emit photons one by one rather than in bursts.
Antibunching is a distinctively quantum phenomenon, implying reduced quantum
fluctuations (squeezing) of light\cite{Reduced_Fluctuations_Zoller_PRL1981} and
sub-Poissonian photon statistics\cite{Sub-Poissonian_Mandel_OL1979}. On the
other hand, it is a very robust effect, exhibited by various fluorophores at
room temperature\cite{Antibunching_dye_room_temperature_Patrick_CPL1997,
Antibunching_Diamonds_Brouri_OL2000, Antibunching_Alivisatos_CPL2000}. We study
the non-classical photon statistics of fluorescence in connection to
fluorescence microscopy and show that it can be used for superresolved
microscopic imaging under realistic conditions.
%We start by reviewing the relevant statistical properties of non-resonant
%fluorescence, then consider the implications for microscopy resolution and,
%finally, discuss possible practical realizations of our scheme.

%\section{second order}
For simplicity, we focus on the case of pulsed excitation, with the pulse duration much shorter, and the interval between pulses much longer than the fluorescence lifetime.
Upon excitation, a single fluorophore emits at most one photon, with a probability p. Such behavior is profoundly different from the Poissonian statistics of classical light. In particular, the variance of the number of fluorescent photons emitted in a series of M excitation cycles is $V=M p (1-p)$, while the mean photon number is $\langle N \rangle = M p$. A classical light source with the same photon flux would yield a variance $V_P = \langle N\rangle$. The variance of the fluorescent photon number is thus reduced by a factor of $(1-p)$ with respect to the classical shot noise.

The nonclassical statistics of fluorescent light can be used to produce superresolved images of fluorophores. Consider a  fluorescent emitter, imaged by a microscopic imaging system onto a pixelated (or scanning) detector with a photon number resolving capability. At every detector position $x$, the probability $P$ to detect a photon emitted by the fluorophore is given by
\begin{equation}\label{P}
P(x)=p Q S h(x-x_0),
\end{equation}
where $Q$ is the quantum efficiency of the detector, $S$ denotes optical
collection efficiency, $h$ is the point spread function of the imaging system
and $x_0$ describes the coordinates of the fluorophore image.  The photon
number variance becomes a function of detector position: $V(x) = M P(x)
\left(1-P(x)\right)$. For a set of several fluorophores, since the emission
events in different fluorophores are uncorrelated, the variance of the total
photon number is given by the sum of variances (\ref{Vx}) for every emitter:
\begin{equation}\label{V}
V(x) = M \sum_\alpha  P_\alpha(x)\left(1-P_\alpha(x)\right) = %\\ =M Q S
%\sum_\alpha p_\alpha h(x-x_\alpha) (1  - QS p_\alpha h(x-x_\alpha)),
\end{equation}
where $p_\alpha$ and $x_\alpha$ are the emission probability and the image position of the fluorophore $\alpha$.

The difference between eq.(\ref{V}) and the classical shot noise variance at
the same mean flux quantifies the degree of antibunching of fluorescent light
and can therefore be called the antibunching signal:
\begin{equation}\label{A}
    A(x) \equiv V(x) - \left\langle N(x) \right\rangle = - M Q^2S^2 \sum_\alpha p^2_\alpha h^2(x-x_\alpha).
\end{equation}
This signal corresponds to an effective point spread function $h_A(x) =
h^2(x)$. In Fourier domain, $h_A$ spans the interval of frequencies twice as
large as that of $h$. The antibunching microscopy thus enables imaging with up
to double resolution, similarly to the resolution improvement attainable with
2-photon microscopy.

\begin{figure}
\includegraphics[width=86mm]{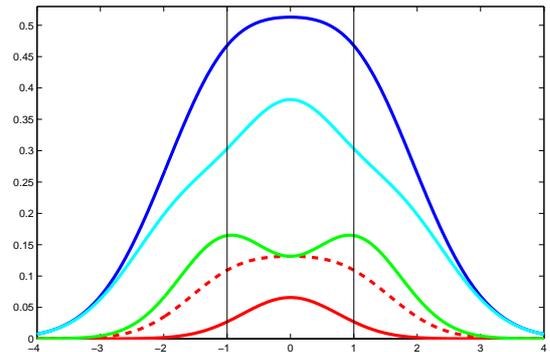}
\caption{Antibunching imaging of two fluorescent emitters. The vertical lines
denote the locations of two fluorescent emitters. The blue line is the regular
fluorescent signal, the cyan line represents the variance given by
eq.(\ref{V}), the green line is the antibunching signal (\ref{A}). The solid
red line shows the probability $F_2$ of a two-photon coincidence event with a
sharp peak in the center. The dashed red line shows the same probability for
two classical emitters, featuring a wider peak. The plots are calculated for
$P_1=P_2=0.4$, with a gaussian point spread function with an RMS width of
$\sigma=1.06$.}\label{two_dots}
\end{figure}

The mechanism of the antibunching microscopy is illustrated in
Fig.~\ref{two_dots} for the case of two identical emitters. The emitters are
not resolved in the fluorescent signal, while two separate peaks are visible in
the antibunching signal.

It is instructive to consider photon statistics in the limit of small photon
flux. Let $F_1(x)$ and $F_2(x)$ be the probabilities of detecting exactly one
and two photons, respectively, at a given detector position. In terms of these
probabilities, the average photon number and the variance become $\left\langle
N \right\rangle = M \left(F_1(x)+2 F_2(x)\right)$ and $V(x) = M \left(F_1+4
F_2-F_1^2\right)$. The antibunching signal (\ref{A}) then takes the form of
$A(x)= M \left( 2 F_2(x) - F_1^2(x)\right)$. Since this expression vanishes for
Poissonian statistics, the antibunching signal can be regarded as a measure of
the lack of two photon coincidence events with respect to classical light.

This observation elucidates the mechanism of the resolution increase shown in
Fig.~\ref{two_dots}: in this example, a coincidence event involves detection of
one photon from each of the fluorophores. The probability of a coincidence
event therefore has a sharp maximum positioned between the two emitters. This
is in contrast to the case of two classical emitters, for which a pair of
photons could as well originate from a single emitter, making the maximum less
sharp.

%\subsection{Covariance}
The antibunching signal (\ref{A}) is determined by the optical signal
autocorrelation at a given detector position. Fluorescence antibunching is also
manifest in the cross correlation between the photon numbers $N_1$, $N_2$
detected in a pair of proximate detectors. Similarly to variance, the
covariance of the two signals $V^\times(x_1,x_2)=\left\langle N_1
N_2\right\rangle - \left\langle N_1\right\rangle\left\langle  N_2\right\rangle$
is a sum of individual fluorophore contributions. For classical light, the
signals observed in the detectors are uncorrelated and therefore covariance
vanishes. In contrast, an individual fluorophore produces only one photon at a
time, which can be detected in only one of the detectors, leading to $ N_1 N_2
= 0$. The covariance antibunching signal can thus be defined as:
\begin{equation}\label{Across}
    A^\times(x_1,x_2) = \sum V_\alpha^\times(x_1,x_2) = - M \sum P^{(1)}_\alpha P^{(2)}_\alpha
    %\\=- M Q^2 S^2 \sum_\alpha  p^2_\alpha h(x_1-x_\alpha) h(x_2-x_\alpha),
\end{equation}
where $x_{1}$,$x_2$ are the detector positions, $P^{(1)}_\alpha$ and
$P^{(2)}_\alpha$ are the detection probabilities for a given fluorophore at the
two detectors. The antibunching signal (\ref{Across}) contains a product of two
optical point spread functions. As we show below, it can be used to produce
higher resolution images along with the autocorrelation antibunching signal
(\ref{A}).

%\section{Numerics}
We numerically tested the resolution improvement in antibunching microscopy by
performing a Monte Carlo simulation of the antibunching imaging process using a
pixelated detector array. For efficient signal utilization, we used both the
autocorrelation antibunching signal of Eq.~(\ref{A}) and the cross correlation
data (\ref{Across}) to form the image. The autocorrelation antibunching signal
at a given pixel, numbered by a two-dimensional index $j$, is given by
\begin{equation}\label{A_num}
A_j = \sum_\alpha \left(p_\alpha^j\right)^2 = V_j - \langle N_j\rangle,
\end{equation}
where $\langle N_j\rangle$ is the mean number of photons, and $V_j$ is the
variance of the photon number. The cross correlation contribution to the second
order antibunching signal was calculated as a weighted sum of the cross
correlations of the pairs of pixels $j\pm\delta$, centered at the pixel $j$.
%The cross correlation decays with $\delta$ at a
%characteristic scale of the width of the point spread function of the imaging
%system. The signal is therefore  maximized by applying a
The simulation was carried out as follows: a stack of frames was generated; in
each frame, for every fluorophore it was decided at random whether it emits a
photon (with a probability $p=0.5$). If a photon was emitted, it was randomly
positioned with a probability density corresponding to a Gaussian point spread
function. The resulting stack of frames was used to compute the second order
antibunching signal according to the following formula:
\begin{equation}\label{discrete}
A_j = \sum_\delta W(\delta) \; \Bigl(\left\langle N_{j+\delta}
N_{j-\delta} \right\rangle - \langle N_{j+\delta}\rangle \; \langle
N_{j-\delta}\rangle\Bigr) - \langle N_j\rangle,
\end{equation}
where $N_j$ is the number of photons detected in the pixel $j$, $\delta$ is a
summation index labeling the pixel pairs, $W(\delta)$ is the weight assigned to
the pixel pairs ($W(0)=1$), and the angular brackets denote averaging over the
set of frames.
%The last term in Eq.~(\ref{discrete}), together with the term in
%the sum with $\delta=0$, compose the contribution of the autocorrelation
%(\ref{A_num}) to the total antibunching signal. The terms with nonzero $\delta$ in
%the sum in Eq.~(\ref{discrete}) represent the cross correlation contribution.
The above analysis does not fully utilize the cross-correlation information:
indeed, only a half of all pixel pairs are centered in a certain pixel. A pair
of, for example, two adjacent pixels has its center between the two pixels. It
is therefore possible to compute the antibunching signal in `virtual pixels' in
between adjacent pixels, i.e. with at least one of the two components of $j$
being half-integer\cite{SOFI_Dertinger_OpEx_2010}. The effective number of
pixels in each direction is thus doubled, increasing the total amount of pixels
by a factor of four. The signal in the `virtual' pixels was calculated using
Eq.~(\ref{discrete}), with the last term omitted and with one or both
components of the summation index $\delta$ assuming half-integer values (so
that $i\pm\delta$ are integer).
\begin{figure}
 \includegraphics[width=100mm]{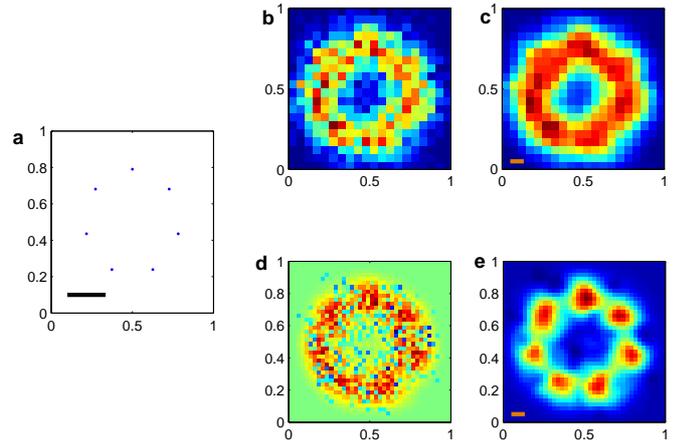}
\caption{A simulation of the second order antibunching imaging: resolving
individual emitters. \textbf{(a)}: seven emitters are arranged in a ring. The
scalebar shows full width at half maximum (FWHM) of the optical point spread
function. \textbf{(b)}: Regular imaging shows a ring structure, but the
individual emitters are not resolved. \textbf{(c)}: the same image as in (b),
smoothed by a Gaussian filter to reduce the noise (filter FWHM is shown by the
scalebar). \textbf{(d)}: the second order antibunching image. \textbf{(e)} the
same image smoothed using the same filter as in (c). The individual emitters
are clearly discerned. The image was formed from $1000$ frames.}\label{ring7}
\end{figure}
The results of the simulation shown in Fig.~\ref{ring7} demonstrate a
significant improvement of resolution: the individual emitters, which cannot be
discerned by regular imaging, are clearly resolved in the antibunching image.

%\section{higher orders}
The antibunching signal, defined above in terms of second order momenta, serves
as a measure of the lack of two-photon coincedence events, hence it can be
called the second order antibunching signal. The n-th order antibunching signal
$A_n(x)$, quantifying the lack of n-photon events, can be defined in terms of
the irreducible parts of the n-th order momenta known as
cumulants\cite{Statistics_Kendall1977}. The defining property of cumulants is
that they are additive for independent random variables, which allows one to
express the cumulants of the observed signal as a sum of the individual
fluorophore contributions. For a single fluorophore, the n-th order cumulant is
given by $C_n=M P(1-P)\,...\,\left(1-(n-1)P\right)$. The cumulants $C_n(x)$ of
the total signal are therefore given by
\begin{equation}\label{Cm}
C_n(x) = M \sum_\alpha P_\alpha(x) \left(1-P_\alpha(x)\right)\,...\, \left(1-(n-1)
P_\alpha(x) \right).
\end{equation}

The antibunching signal of order $n$ can then be defined as
\begin{equation}\label{An}
    A_n=M \sum P_\alpha^n,
\end{equation}
which can be expressed via cumulants of order $k \leq n$ using Eq.~(\ref{Cm}):
\begin{equation}\label{Im}
A_n(x) =  M \sum_k  (\hat{R}^n)_{1 k}\; C_k(x),
\end{equation}
where $(\hat{R}^n)_{1 k}$ is the first row of the n-th power of a matrix
$\hat{R}$, all elements of which are zero except
\begin{equation}\label{R}
R_{k k}=-R_{k-1 k} = 1/k.\end{equation} The signals $A_n(x)$ vanish in
the classical limit, and are therefore a valid local measure of the degree
of antibunching. For $n=2$ the above expressions yield the second order antibunching signal described above. Substituting eq.~(\ref{P}) into the definition (\ref{An})  one obtains
\begin{equation}\label{Ihm}
A_n(x) = M Q^n S^n \sum_\alpha p_\alpha^n
\left[h(x-x_i)\right]^n.\end{equation} Effectively, the n-th order antibunching
signal corresponds to a point spread function $h_m(x) = \left[h(x)\right]^m$.
Similarly to the second order, this enables imaging with resolution up to n
times better than diffraction limited in three
dimensions\cite{Gustafsson_SSI_PNAS05, Pupil_Filters_Schwartz_OL2009}.

%Similarly to the second order, to utilize the signal efficiently it is
%necessary to consider cross correlations between several detectors. The
%higher-order generalizations of covariance are n-detector joint cumulants.
%In classical optics, the photon number distributions in different
%detectors are uncorrelated, and thus all joint cumulants turn to zero. In
%contrast, a fluorophore can only send a photon to one detector at a time, which
%brings about negative correlations between the detectors:
%\begin{equation}\label{cross}
%   C^{\times}_n = - M \sum_\alpha P_\alpha^{(1)} P_\alpha^{(2)} ... P_\alpha^{(m)},
%\end{equation}
%where $P_\alpha^k$ is the probability for k-th detector to detect a photon emitted by fluorophore $\alpha$.

%\section{discussion}
%An important limitation of the present superresolution scheme is that it cannot
%be used to increase the resolution of a low aperture imaging system: the
%performance of this method depends critically on the detection efficiency,
%which includes the geometrical collection efficiency $S$.The scheme can thus
%be effectively realized only with high numerical aperture optics.
Quantum fluorescence imaging requires fluorophores with as high quantum yield
as possible. Fortunately, many of the fluorescent markers widely used in
bio-imaging, such as organic dyes, have a quantum yield approaching unity. Very
high quantum yield has also been demonstrated with colloidal semiconductor
quantum dots\cite{Nonblinking_dots_Chen_JACS2008}.

Many fluorescent single photon emitters exhibit random variations of brightness
known as blinking. Blinking increases the observed photon number fluctuations,
and could be expected to change the photon statistics to super-Poissonian. It
turns out, however, not to be the case: indeed, for a single emitter, the
photon number follows a Bernoulli distribution even in the presence of
blinking, the only consequence of which is an effective reduction of the
emission probability. As long as blinking of individual fluorophores is
uncorrelated, the antibunching properties of emission statistics persist for an
arbitrary number of emitters.

The photon number distributions required for computing the cumulants in
Eq.~(\ref{Im}) can be determined with a scanning number resolving
detector\cite{number_resolving_Lita_OpEx2008,
N_photon_detection_Shapiro_PRL2010}. Another option is using a regular or an
electron multiplying charged coupled device (ccd)\cite{Centroids_PRL2009}. An
electron multiplying ccd in the photon counting regime can be used as a
pixelated number resolving detector, provided that the number of pixels
sufficiently large so that the probability of detecting more than one photon in
a pixel is small. Interestingly, a regular ccd, which is naturally a photon
number resolving detector, can also be used despite signifant amount of noise
in the output. Since the noise is statistically independent from the number of
photons, it can  be removed from the signal by subtracting the pre-measured
cumulants of the noise from the observed cumulants.

The antibunching microscopy can be regarded as a quantum version of
superresolution optical fluctuation imaging
(SOFI)\cite{SOFI_DertingerPNAS2009}: in this technique, the n-th order signal
is given by n-th cumulant, without the lower order `correction' terms appearing
in Eq.~(\ref{Im}). Interestingly, the antibunching signal vanishes in the
classical limit, instead of turning into the corresponding SOFI expression.
This is the case because the two schemes exploit different sources of
non-Poissonian statistics: while SOFI quantifies the super-Poissonian
brightness fluctuations of essentially classical sources, in the present scheme
the signal is due to the reduction of the quantum fluctuations with respect to
the classical shot noise. The antibunching signal is thus generated by steadily
emitting fluorophores, which enables continuous superresolved monitoring of the
samples stained with fluorescent markers.

%\section{conclusion}
In conclusion, we propose a fluorescence microscopy imaging modality
that allows for sub-diffraction-limited imaging by virtue of quantum properties
of fluorescence emission. Despite being ostensibly quantum, the technique does
not require a non-classical light source and does not depend on fragile quantum
interference effects. The proposed method can be implemented with current
technology, or indeed with a regular fluorescence microscope.
%\bibliographystyle{apsrev}
%\bibliography{microscopy,quantum_microscopy}

\end{document}